\documentclass[a4paper,11pt]{article}
\usepackage{pos}

\title{Performance of LACT Array: Instrument Response Functions and Source Prospects}

\author*[a]{Zhipeng Zhang}
\author[a,b]{Ruizhi Yang}
\author[b,c]{Shoushan Zhang}

\affiliation[a]{University of Science and Technology of China,\\ No.96 JinZhai Road, Baohe District, Hefei, Anhui 230026, China}

\affiliation[b]{TIANFU Cosmic Ray Research Center,\\ Chengdu, Sichuan 610213, China}

\affiliation[c]{Institute of High Energy Physics, Chinese Academy of Sciences,\\ 19B Yuquan Road, Shijingshan District, Beijing 100049, China}

\emailAdd{zhipzhang@mail.ustc.edu.cn}
\emailAdd{yangrz@ustc.edu.cn}

\abstract{Large Array of imaging atmospheric Cherenkov Telescope (LACT) is an array of 32 Cherenkov telescopes with 6-meter diameter mirrors to be constructed at the LHAASO site, aiming to enhance our understanding of ultra-high energy gamma ray astronomy. This work presents a detailed performance assessment of the LACT array, focusing on the IRFs for both an 8-telescope subarray configuration optimized for large zenith angle observations (60°) and the full 32-telescope array, with a particular emphasis on a 20° zenith angle configuration for lower energy threshold observations.

We have generated IRFs using extensive Monte Carlo simulations of gamma-ray showers and the detector response. The IRFs include the effective area, angular resolution, and energy resolution as a function of reconstructed energy and offset angle. Crucially, these IRFs are produced in the standard Data format for Gamma ray astronomy (GADF), ensuring interoperability with existing analysis tools like Gammapy and ctools and enabling seamless integration into scientific workflows.

In this work, we also have used these GADF-format IRFs to simulate observations of key astrophysical sources and assess the LACT array's capabilities for morphology studies and spectral analysis.}

\ConferenceLogo{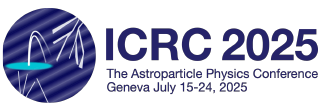}

\FullConference{39th International Cosmic Ray Conference (ICRC2025)\\
 15–24 July 2025\\
Geneva, Switzerland\\}

\begin{document}
\maketitle

\section{Introduction to LACT Project}
In early 2024, LHAASO released its first catalog, which included 90 very-high-energy gamma-ray sources and, notably, 43 ultra-high-energy (UHE) gamma-ray sources, thereby dramatically expanding the landscape of TeV/PeV astronomy\cite{LHAASO-first}. Leveraging its exceptional sensitivity, LHAASO has identified numerous candidate PeVatrons, such as pulsar wind nebulae (PWNe), young massive clusters (YMCs), supernova remnants (SNRs), and microquasars\cite{Ultra-High}. Despite these advances, the limited angular resolution of LHAASO hinders detailed study of individual sources and the precise localization of particle acceleration regions. To overcome these limitations and further advance in this direction, the LACT project has been initiated.

Unlike LHAASO, imaging atmospheric Cherenkov telescopes (IACTs) operate only during moonless nights, resulting in a much lower duty cycle. As a result, to observe a comparable number of ultra-high-energy gamma-ray events, LACT requires an effective area significantly larger than that of current-generation IACTs—exceeding $1 ~\rm km^2$ in order to compensate for the reduced observation time.

\begin{figure}[htbp]
    \centering
    \begin{minipage}[t]{0.48\textwidth}
        \centering
        \includegraphics[width=\textwidth]{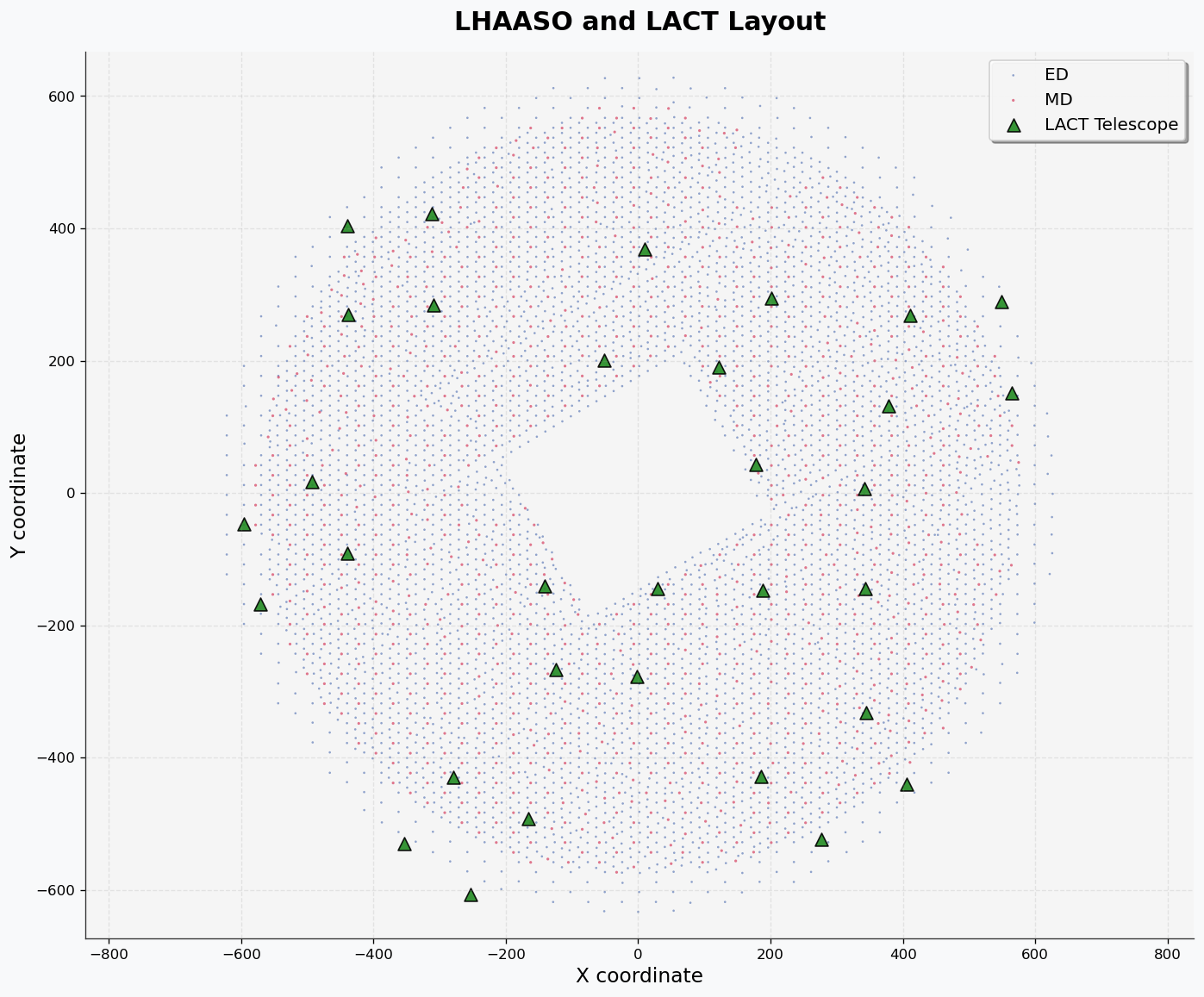}
        \caption{Layout of LACT and LHAASO}
        \label{fig:layout-lact}
    \end{minipage}
    \hfill
    \begin{minipage}[t]{0.48\textwidth}
        \centering
        \includegraphics[width=\textwidth]{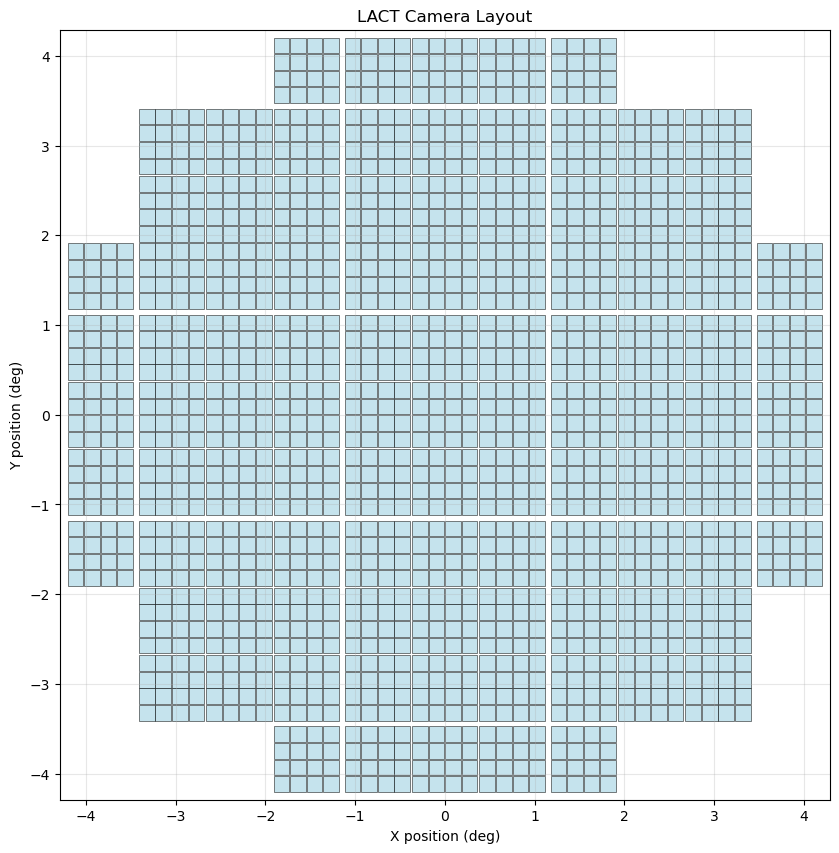}
        \caption{Camera layout of LACT}
        \label{fig:camera-lact}
    \end{minipage}
\end{figure}

To achieve a large effective area, LACT is composed of 32 telescopes and supports two complementary observation modes:
\begin{enumerate}
\item \textbf{Normal Observation Mode:} In this mode, LACT operates as a unified array, delivering excellent performance across a wide energy range—from a few hundred GeV up to hundreds of TeV. In addition to enabling the detection of ultra-high-energy gamma rays, this configuration also allows for the obsservation of variable and transient sources, such as active galactic nuclei (AGN) and gamma-ray bursts (GRBs).
\item \textbf{Large Zenith Angle Observation Mode:} By observing cosmic gamma rays at large zenith angles, IACTs can  significantly increase its effective area\cite{MAGIC-LZ} \cite{VERITAS-LZ}. This mode is especially suited for detecting ultra-high-energy gamma-ray events.
\end{enumerate}
To fully exploit the advantages offered by large zenith angle observations, we have conducted detailed Monte Carlo simulations to optimize the layout of LACT\cite{LACT-CPC}. After conducting on-site investigations at the LHAASO site, the final layout of the LACT array has been determined, as shown in Figure~\ref{fig:layout-lact}. Each 6-meter LACT telescope is equipped with a camera providing a wide field of view exceeding \(8^\circ\), with individual pixels covering an angular size of approximately \(0.19^\circ\). The layout of the LACT camera is shown in Figure~\ref{fig:camera-lact}. 

\section{The Peformance of LACT}
To evaluate the array’s performance in the two aforementioned observation modes, we performed detailed Monte Carlo simulations based on the actual layout. These simulations included on-axis point gamma rays, diffuse gamma rays, and diffuse proton events.The detail of reconstruction method can be found in \cite{LACT-CPC}. It should be noted, however, that the selection cuts used here have not been fully optimized.
\subsection{20 degree Zenith Angle Performance}
At a zenith angle of $20^{\circ}$, LACT will operate as a full array to maximize its capabilities. Owing to the closed-layout within the cells, the threshold energy of the LACT array can be as low as $250~\mathrm{GeV}$, which is particularly advantageous for the observation of GRBs and AGNs.

\begin{figure}[htbp]
    \centering
    \includegraphics[width=0.6\textwidth]{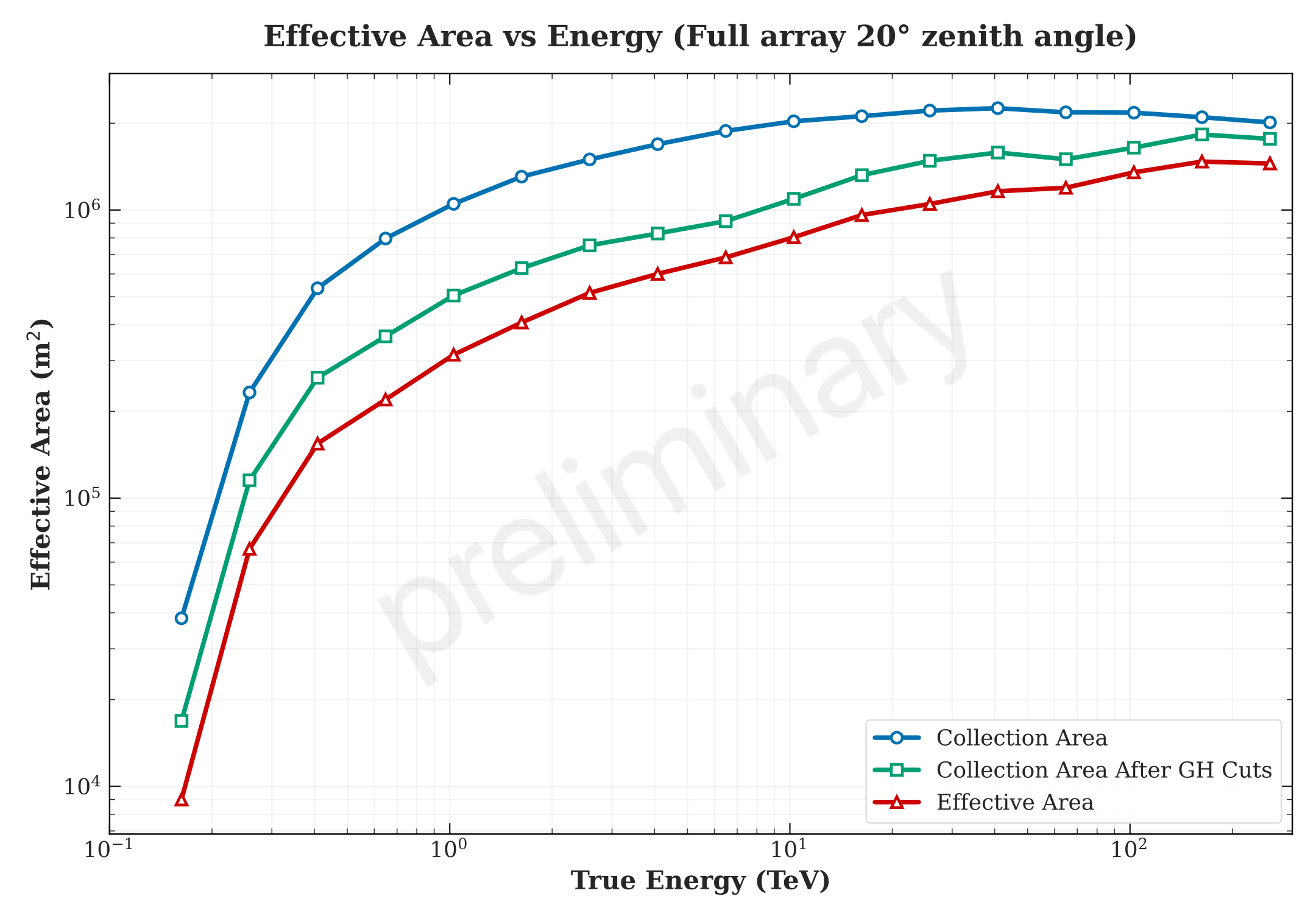}
    \caption{Effective area for the LACT full array at 20$^\circ$.}
    \label{effe_20}
\end{figure}

\begin{figure}[htbp]
    \centering
    \includegraphics[width=0.6\textwidth]{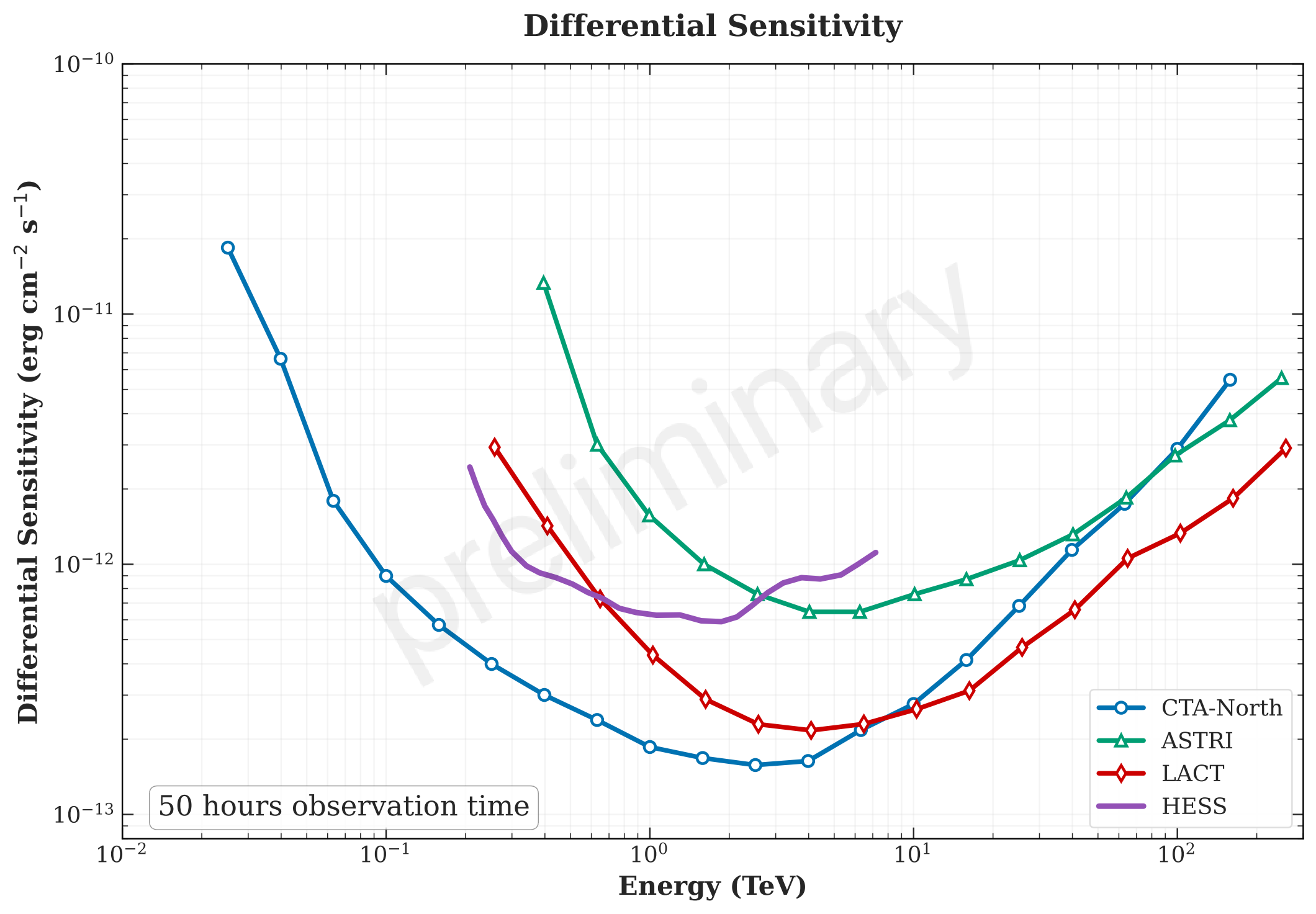}
    \caption{Differential sensitivity for the LACT full array at 20$^\circ$, compared with other instruments in the Northern Hemisphere.}
    \label{sens_20}
\end{figure}

In addition to its capability for detecting extra-galactic sources, the 32 telescopes and large field-of-view of LACT camera enable it to achieve a collection area greater than $2~\mathrm{km}^2$. The effective area of LACT is shown in Figure~\ref{effe_20}. We also compare the sensitivity of the LACT array with that of other instruments located in the Northern Hemisphere, as shown in Figure~\ref{sens_20}. The results indicate that, above $10 ~\rm TeV$, the LACT array will surpass both existing and planned IACTs, making it the most sensitive facility in the northern hemisphere. An interesting ongoing effort involves leveraging LHAASO-KM2A to improve the retention rate of gamma ray events, thereby further enhancing sensitivity at energies above tens of TeV. Moreover, since the collection area of LACT will be twice that of LHAASO-KM2A \cite{KM2A-CPC}, it enables investigating the KM2A response to events that occur outside the array.


\subsection{60 degreee Zenith Angle Performance}
For observations at low zenith angles, the high altitude of LACT (\(4{,}400\,\mathrm{m}\)) results in more pronounced image leakage compared to lower-altitude sites. This increased leakage significantly degrades the instrument’s performance, particularly at energies above several tens of TeV. To address this challenge and optimize sensitivity for ultra-high-energy gamma rays (above \(100\,\mathrm{TeV}\)), We propose conducting observations at larger zenith angles. As we demonstrate below, this approach not only increases the effective area but also, by dividing the array into four subarrays, potentially allows for up to a fourfold increase in observation time. 
\begin{figure}[htbp]
    \centering
    \includegraphics[width=0.6\textwidth]{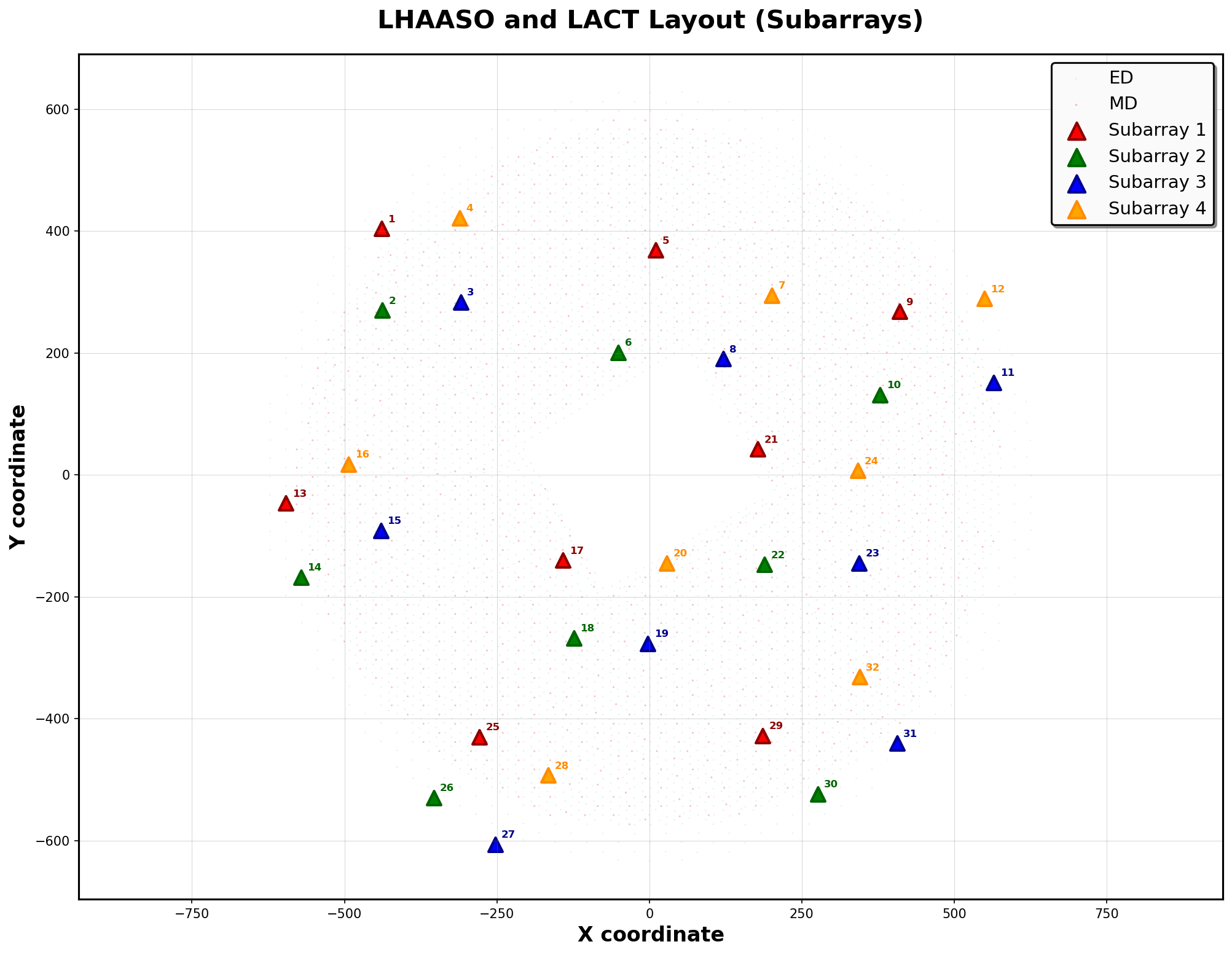}
    \caption{Layout of LACT array, different color represent different subarrays in the large zenith angle observation.}
    \label{fig:lay_with_sub}
\end{figure}

\begin{figure}[htbp]
    \centering
    \begin{minipage}[t]{0.43\textwidth}
        \centering
        \includegraphics[width=\textwidth]{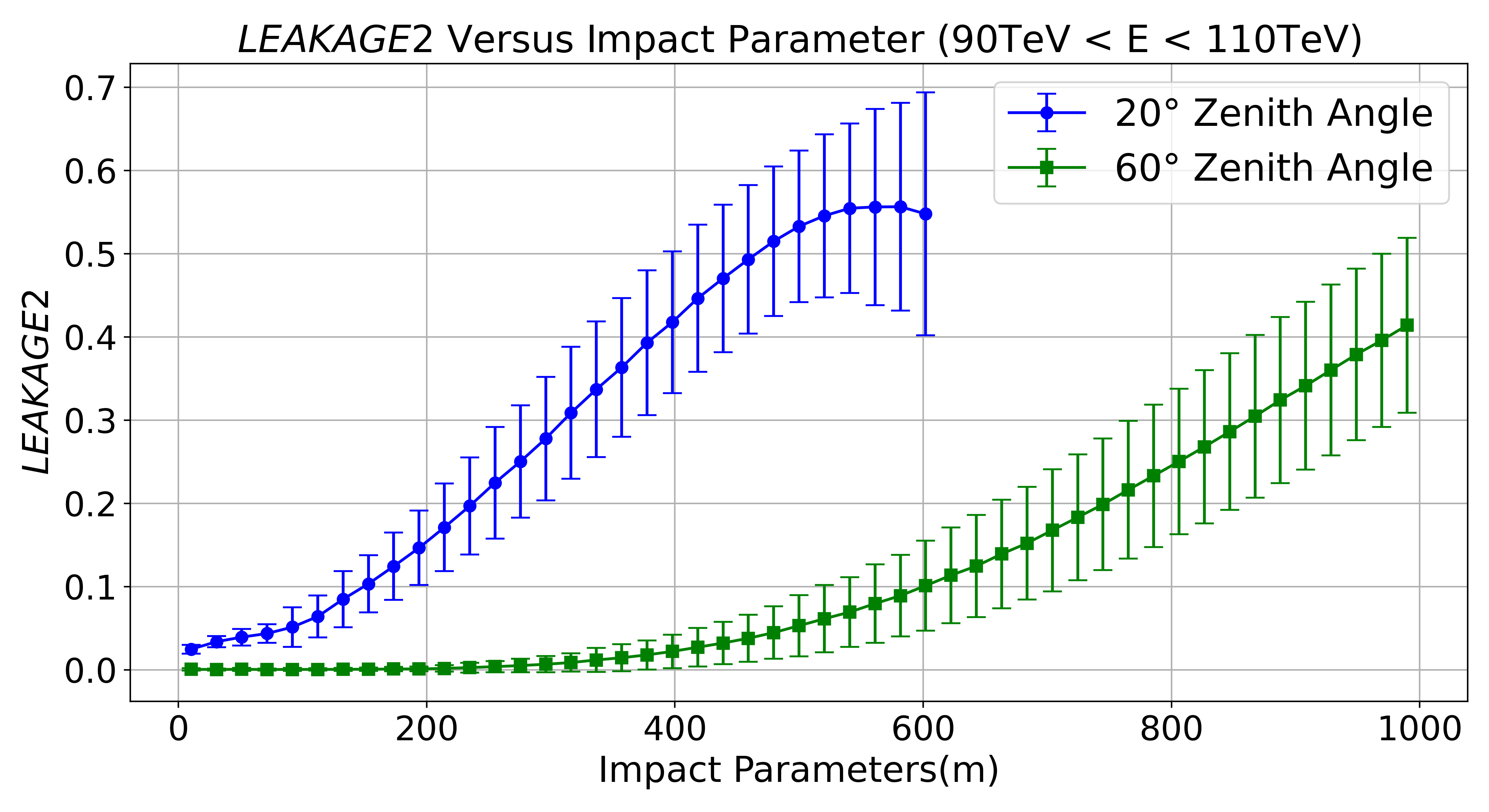}
        \caption{Leakage as a function of impact parameter. The error bars represent the standard deviation of the leakage distribution within each bin.}
        \label{fig:leakage}
    \end{minipage}
    \hfill
    \begin{minipage}[t]{0.48\textwidth}
        \centering
        \includegraphics[width=\textwidth]{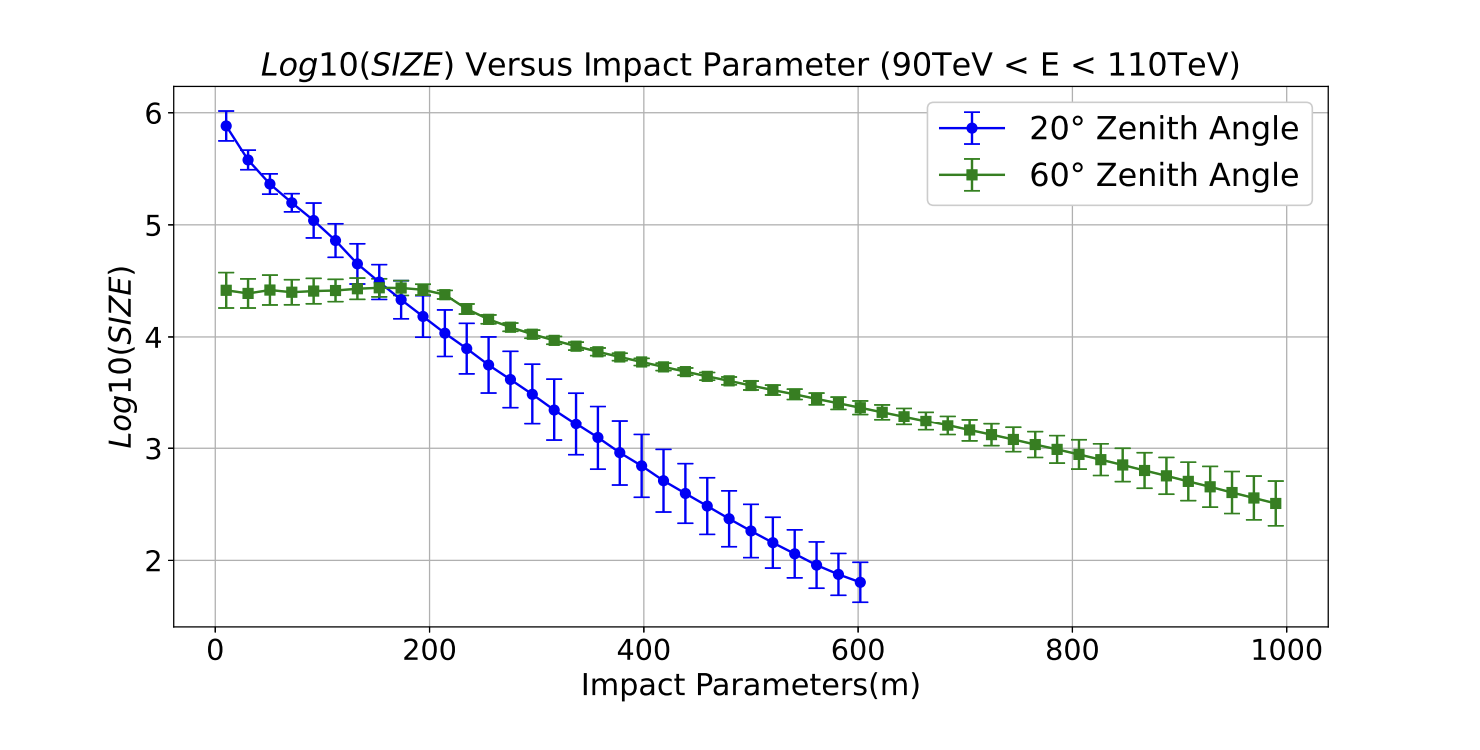}
        \caption{Lateral distribution of photon density. The error bars here is the standard deviation of the intensity within each bin.}
        \label{fig:lateral}
    \end{minipage}
\end{figure}

As illustrated in Figure~\ref{fig:leakage} and Figure~\ref{fig:lateral}, observations at large zenith angles exhibit a significantly flatter lateral distribution of photon density and substantially reduced image leakage. These characteristics are crucial for ultra-high-energy gamma-ray detection, because for the larger multiplicity and  better reconstruction. 

\begin{figure}[htbp]
    \centering
    \includegraphics[width=0.6\textwidth]{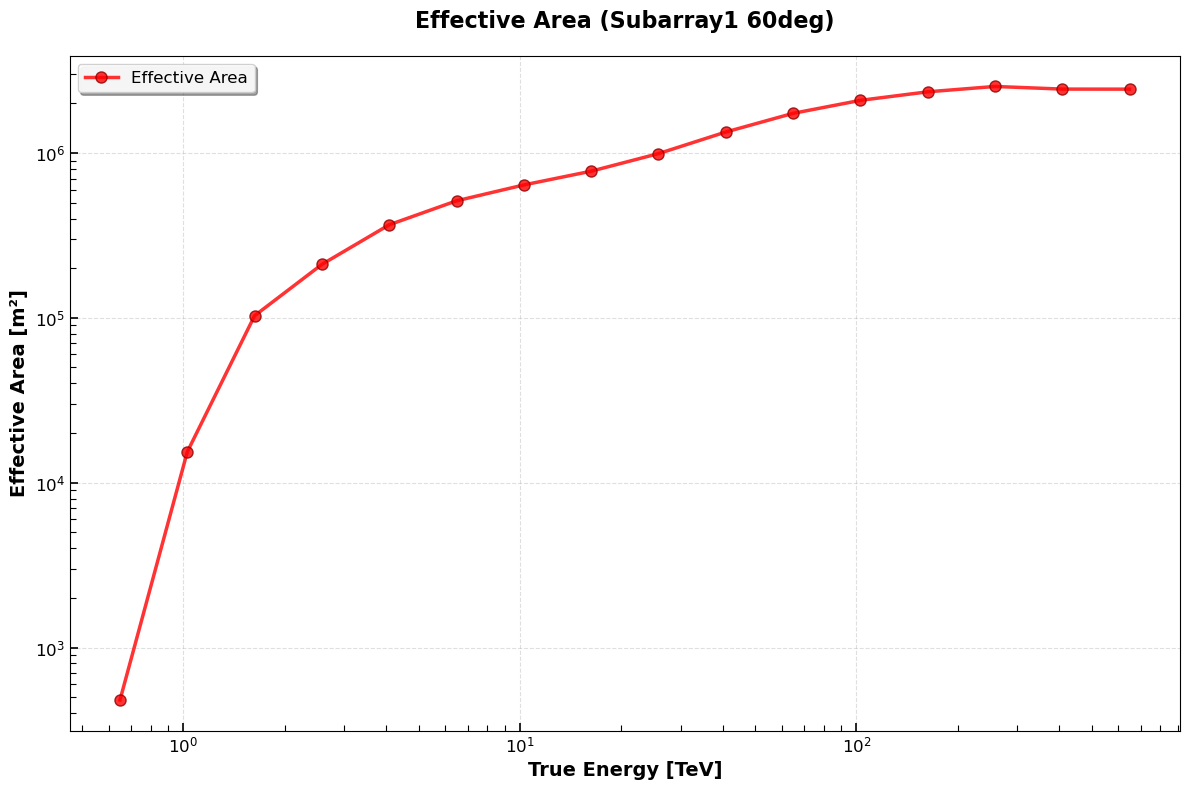}
    \caption{Effective area for the LACT subarray at 60$^\circ$.}
    \label{effe_60}
\end{figure}

\begin{figure}[htbp]
    \centering
    \includegraphics[width=0.6\textwidth]{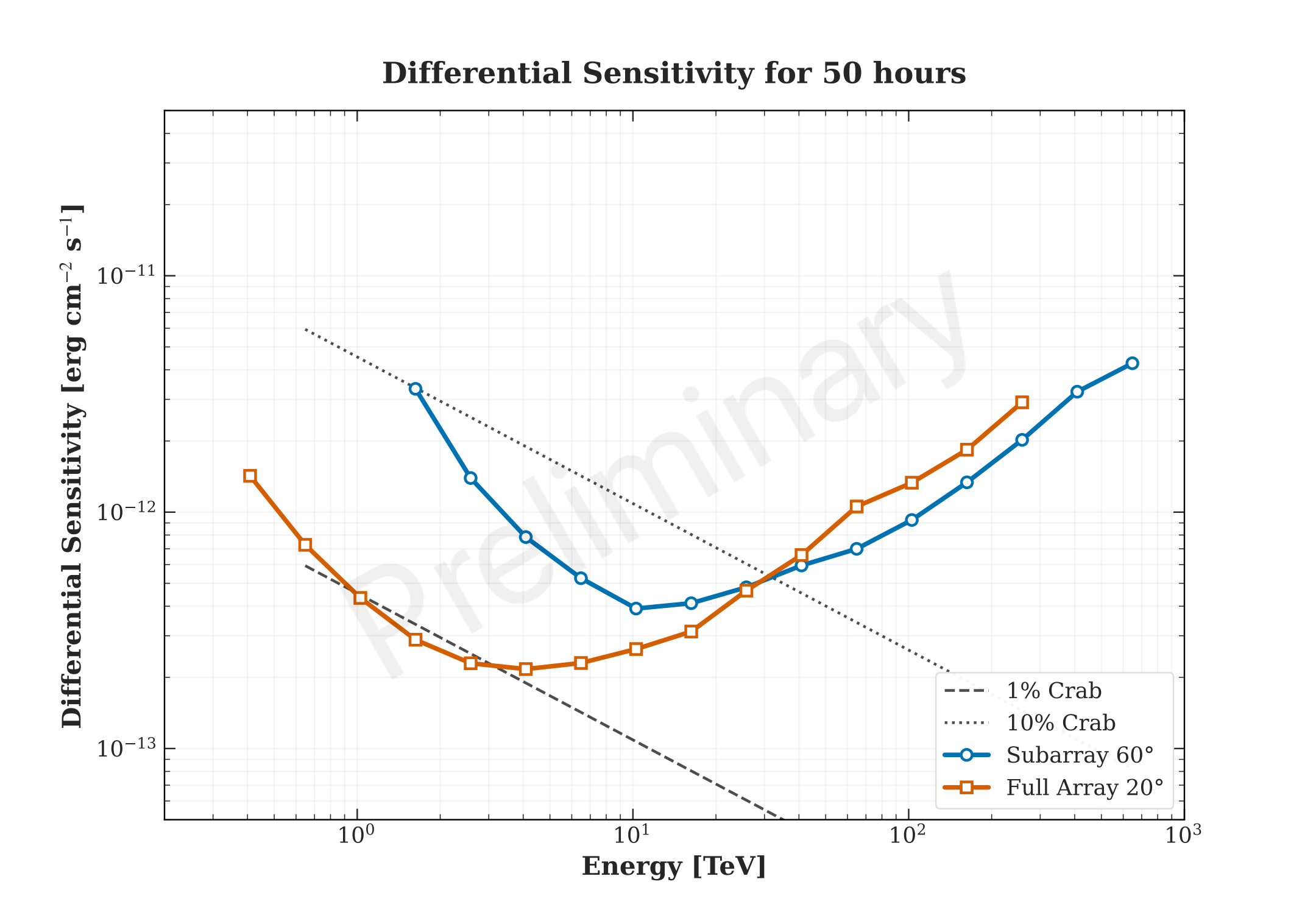}
    \caption{Differential Sensitivty for LACT-subarray1 for 50hours}
    \label{sens_60}
\end{figure}

To mitigate the effects of nearly “parallel images” \cite{HESS-DISP} and to achieve improved stereoscopic observation of showers that occur far from the array center, we divide the 32 telescopes of LACT into four sub-arrays, each comprising eight well-separated telescopes. We then investigate the performance for the subarray. The effective area and sensitivity of this subarray is shown in Figure ~\ref{effe_60} and Figure ~\ref{sens_60}. Although each subarray consists of only eight telescopes, the large zenith angle mode allows each telescope to detect showers up to nearly 800 meters away. As a result, the effective area of a subarray can exceed \(2.5~\mathrm{km}^2\), which is highly advantageous for detecting rare ultra-high-energy photons. Furthermore, compared to the LACT array operating at \(20^\circ\), the subarray demonstrates markedly improved sensitivity above \(30~\mathrm{TeV}\), despite having a somewhat higher energy threshold of approximately \(2~\mathrm{TeV}\). It is also important to note that, since the large zenith angle observation mode comprises four nearly identical subarrays, the total observation time can be increased by a factor of four. This extended observation time is particularly beneficial for IACTs.


\section{Use Gammapy to simulate the observation of Galactic Center Region}
Prior to the construction of  LHAASO, the Galactic center region provided the first hint of a potential Pevatron. Results from HESS indicated that the diffuse emission spectrum of the central molecular zone (CMZ) exhibits no cutoff up to tens of TeV \cite{HESS-GC}. Although the Galactic center region is not accessible to LHAASO, it can be observed by LACT when operating in large zenith angle modes. 

\begin{figure}[htbp]
    \centering
    \includegraphics[width=0.8\textwidth]{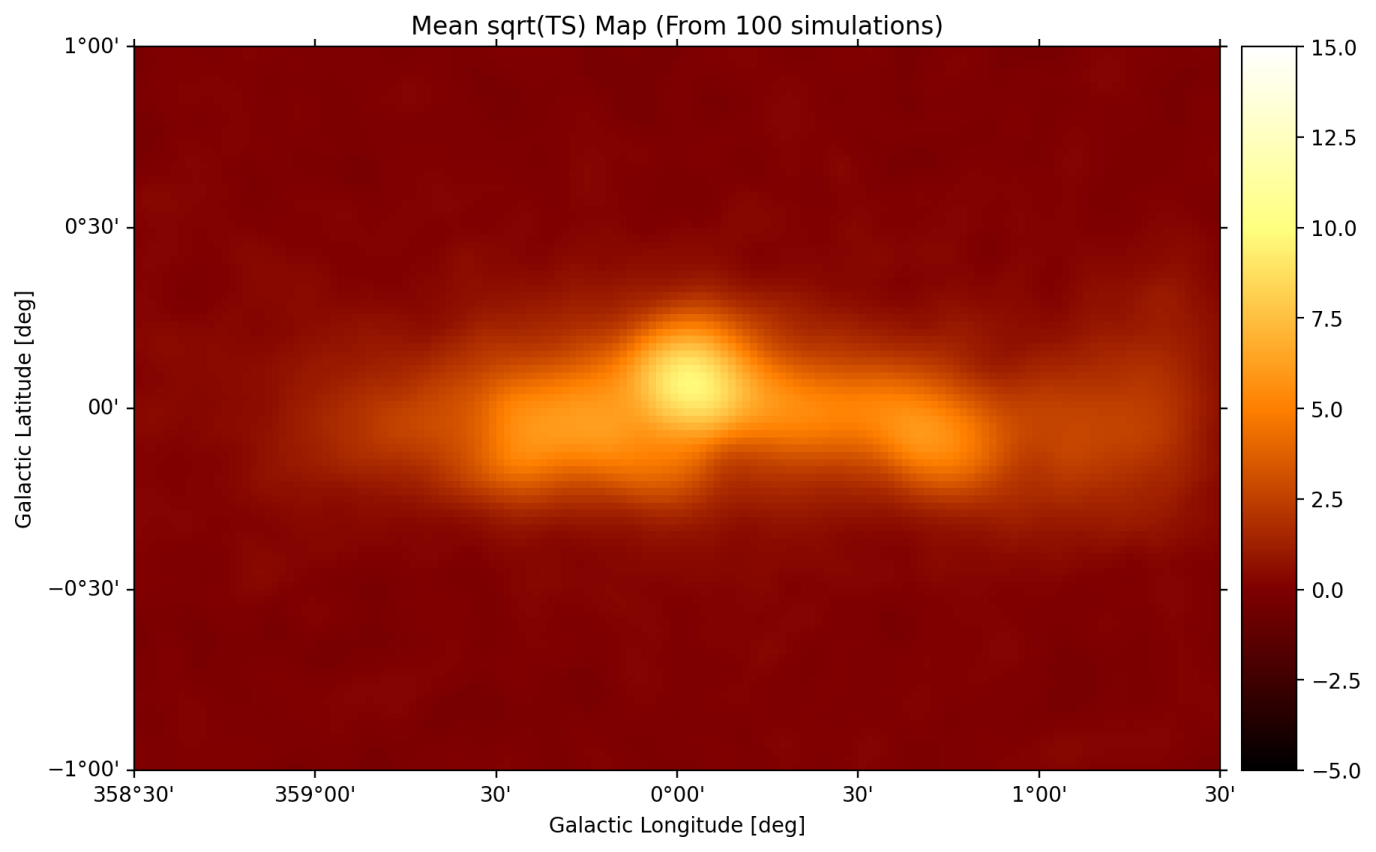}
    \caption{TsMap for 50hour observation towards Galactic center region.}
    \label{TsMap}
\end{figure}

Here, we present preliminary results from simulations of the Galactic center region performed using Gammapy\cite{GAMMAPY}. During the optimal observing season for LACT (approximately October to April), the Galactic center can be observed for more than  50 hours per year. The spatial and spectral models used in our simulation are derived from $\sim 260 ~\rm hours$ of HESS observations \cite{HESS-GC-M}.The averaged TS map from 100 realizations of 50-hour simulated observations is presented in Figure~\ref{TsMap}. As shown, significant diffuse gamma-ray emission from the CMZ is clearly detectable. We also assessed the capability to identify a spectral cutoff: with 100 hours of observation, LACT can robustly detect a cutoff energy up to $80 ~\rm TeV$

\section{Conclusion}
In this contribution, we have presented a comprehensive performance evaluation of the upcoming LACT array, based on detailed Monte Carlo simulations. Our findings highlight LACT's capabilities in two distinct but complementary observation modes.

The full 32-telescope array, operating at a $20^{\circ}$ zenith angle, will achieve an energy threshold as low as $\sim 250 ~\rm GeV$ and an effective area $\sim 1.6 ~\rm km^2$. This configuration establishes LACT as the most sensitive imaging atmospheric Cherenkov telescope facility in the Northern Hemisphere for energies above 10 TeV, making it ideal for studying both galactic and extra-galactic sources.

The subarray configuration for large zenith angle observations is specifically designed for ultra-high-energy studies. Due to reduced image leakage and a flatter lateral distribution, the subarray achieves a substantially larger effective area and potentially even better angular resolution for showers with energies above several tens of TeV. A key advantage of this mode is the ability to operate the four subarrays independently, thereby potentially quadrupling the total available observation time.

In summary, LACT is poised to be a powerful instrument for the next generation of gamma-ray astronomy. It will provide the detailed morphological and spectral analysis necessary to resolve the PeVatron candidates discovered by LHAASO, thereby playing a pivotal role in advancing our understanding of the ultra-high-energy universe.

\end{document}